\documentstyle[11pt]{article}
\def\ft#1#2{{\scriptstyle {#1 \over #2}}}

\def\ww3{{$W_3$}}

\def\del{\partial}
\def\a{\alpha}

\def\ada{{\alpha\dot\alpha}}
\def\da{{\dot\alpha}}
\def\bdb{{\beta\dot\beta}}
\def\tb{{\tilde\beta}}
\def\t{\theta}

\begin{document}
\topmargin 0pt
\oddsidemargin 5mm
\begin{titlepage}
\begin{flushright}
CTP TAMU-62/94\\
SISSA-175/94/EP\\
hep-th/9411101\\
\end{flushright}
\vspace{1.5truecm}
\begin{center}
{\bf {\Large BRST Quantisation of the $N=2$ String in a Real Spacetime
Structure}}
\vspace{1.5truecm}

{H. L\"u\footnote{Supported in part by the
U.S. Department of Energy, under grant DE-FG05-91-ER40633}\ and\  C.N.
Pope$^{1,}$\footnote{Supported in part by the EC Human Capital and Mobility
Programme, under contract number ERBCHBGCT920176.}
}
\vspace{1.1truecm}

{\small Center for Theoretical Physics, Texas A\&M University,
                College Station, TX 77843-4242} \vspace{1.1truecm}


\end{center}
\vspace{1.0truecm}

\begin{abstract}
\vspace{1.0truecm}

      We study the $N=2$ string with a real structure on the $(2,2)$
spacetime, using BRST methods.  Several new features emerge.  In the
diagonal basis, the operator $\exp(\lambda \int^z J^{\rm tot})$, which is
associated with the moduli for the $U(1)$ gauge field on the world-sheet, is
local and it relates the physical operators in the NS and R sectors.  However,
the picture-changing operators are non-invertible in this case, and physical
operators in different pictures cannot be identified.   The three-point
interactions of all physical operators leads to three different types of
amplitudes, which can be effectively described by the interactions of a
scalar NS operator and a bosonic spinorial R operator.  In the off-diagonal
bases for the fermionic currents, the picture-changing operators are
invertible, and hence physical operators in different pictures can be
identified. However, now there is no local operator $\exp(\lambda \int^z
J^{\rm tot})$ that relates the physical operators in different sectors.  The
physical spectrum is thus described by one scalar NS operator and one spinorial
R operator.    The NS and R operators give rise to different types of
three-point amplitudes, and thus cannot be identified.

\end{abstract}
\end{titlepage}
\newpage
\pagestyle{plain}
\section{Introduction}

\indent The $N=2$ string acquired a new physical interpretation with
the work of Ooguri and Vafa \cite{OV}.  The picture that emerged was that
there was just one physical state in the theory, which corresponds to a
self-dual graviton in the closed string and to a self-dual Yang-Mills field
in the open string.  The analysis was carried out in a traditional approach
which, in BRST language, corresponds to restricting attention to physical
states of standard ghost structure.   However, recent experience with
certain string theories has shown that the complete description of the
physical spectrum may require the inclusion of excitations of the ghost
fields.  Thus it is worthwhile to study the $N=2$ string in the BRST
framework. This subject has already received some attention in the
literature, both before and after the appearance of Ref.~\cite{OV}. Although
incomplete, the results of these investigations seemed to indicate that a
full BRST treatment led to nothing new.

     The fermionic constraints of the $N=2$ string necessarily break the
full Lorentz covariance of the $(2,2)$ signature target spacetime.  Owing to
the fact that $SO(2,2)$ is isomorphic to $SL(2,R)_{\rm L}\times SL(2,R)_{\rm
R}$, one may use a two-component spinor notation in which a spacetime vector
index $\mu$ is replaced by the pair of two-component $SL(2,R)_{\rm R}$ and
$SL(2,R)_{\rm L}$ indices $(\alpha\dot\alpha)$. In this notation, it is the
covariance under $SL(2,R)_{\rm L}$ that is broken.  There are three ways of
choosing the mapping between the $\mu$ and $(\alpha\dot\alpha)$ indices, one
of which corresponds to grouping the two timelike directions into one
complex coordinate, and the two spacelike directions into another complex
coordinate.  This choice corresponds to taking a {\it complex} structure on
the spacetime, and is the one that is usually considered for the $N=2$
string.  The other two possibilities are essentially equivalent to each
other, and correspond instead to grouping timelike and spacelike directions
together in pairs.  This corresponds to taking a {\it real} structure on the
spacetime.  It is this choice that we adopt in the present paper.

    New features are highlighted when one chooses to study the $N=2$
string in the real structure on the $(2,2)$ spacetime.  This is because the
momentum components $p^{\alpha\dot1}$ and $p^{\alpha\dot 2}$ of
$p^{\alpha\dot\alpha}$, which would be complex conjugates of each other for
the complex structure on spacetime, are instead independent in the real
structure.  This means that special cases can arise when one vanishes and
the other does not.

      In this paper, we shall examine some aspects of the BRST quantisation
of the $N=2$ string with a real spacetime structure.  A crucial feature of
the $N=2$ string that is not present for strings with $N\le 1$ is that the
bosonisations of the ghost fields for the fermionic currents in different
bases are not locally related to one another.   Thus one may expect that the
physical spectrum might be different for different basis choices.  In this
paper, we study the physical spectrum of the $N=2$ string in various bases,
and we find that indeed there are differences, and we compare the results in
different bases with those of Ref.~\cite{OV}.   A key point is that in the
diagonal basis of fermionic currents the picture-changing operators that
normally establish equivalences between classes of states at different
picture numbers become non-invertible.  As a consequence of this, it turns
out that infinitely many additional massless physical states arise in the
spectrum in such bases. These are states which, for generic on-shell
momenta, are related by the picture-changing operators, but which are no
longer related when certain components of their momenta vanish.  This
problem is more acute when one uses a real structure in spacetime, for the
reasons discussed in the previous paragraph.

     In the diagonal basis, there is a local operator $\exp(\lambda \int^z
J^{\rm tot})$ that relates the physical operators in the Neveu-Schwarz and
Ramond sectors.   It can be understood as the operator whose insertion at
punctures on the sphere describes the residual moduli for the $U(1)$ gauge
field on the world-sheet.   Thus in the diagonal basis, the NS and R sectors
are equivalent after the integration of the $U(1)$ moduli.  However, in this
diagonal basis, the picture-changing operators are non-invertible, and the
physical operators at different pictures cannot be identified.  We obtain
the interactions of all the physical operators.  There are three types of
different momentum-dependant three-point amplitudes.  We shall show that
they can be effectively described by the interactions of a scalar NS
operator and a bosonic spinorial R operator.

      The story in the off-diagonal basis is quite different.  The
picture-changing operators become invertible, and hence the physical
operators in different pictures can be identified.  However, the operator
$\exp(\lambda\int^z J^{\rm tot})$ becomes non-local.  The physical operators
in the NS and R sectors, which are defined in the local Hilbert space, can
no longer be related by the integration of the $U(1)$ moduli.  Thus the
physical spectrum is described by one scalar NS operator and one spinorial R
operator.  The differences between the two operators are emphasised if one
computes the three-point amplitudes.  There are two different three-point
amplitudes, namely $\langle {\rm NS, NS, NS}\rangle$ and $\langle {\rm R, R,
NS}\rangle$.  The former has explicit momentum dependence and will vanish at
some special momenta, whilst the latter is proportional to the inner product
of the two polarisation spinors.

      The paper is organised as follows.  In sections 2 and 3, we derive
the NS spectrum of the $N=2$ string in various bases for the fermionic
currents.   In section 4, we investigate the interactions for the physical
operators in the diagonal basis.  In section 5, we obtain the interactions
in the off-diagonal basis.   We conclude the paper in section 6.  In the
appendix, we give the results for the inverses of the picture-changing
operators in a generic off-diagonal basis.

\section{$N=2$ String in the Diagonal Basis}
\subsection{The BRST Operator}

\indent   In this section, we describe the $N=2$ string and its physical
spectrum in a basis where the
fermionic currents are $\pm$ eigenstates under the $U(1)$ current.  We consider
a matter system consisting of the four spacetime coordinates $X^\ada =
\sigma_\mu^\ada\, X^\mu$, the two-component Majorana-Weyl spinor $\theta^\a$
and its world-sheet conjugate momentum $p_\a$.  In the language of conformal
field theory, these satisfy the OPEs
\begin{equation}
X^\ada(z)\, X^\bdb(w) \sim -\epsilon^{\alpha\beta} \epsilon^{\da\dot\beta}
\,{\log} (z-w)\ ,\qquad
p_\a(z)\, \theta_\beta(w) \sim {\epsilon_{\alpha\beta}\over (z-w)}\ .
\end{equation}
When we need to be explicit, we use conventions in which the spacetime
metric is given by $\eta_{\mu\nu} ={\rm diag}(-1,-1,1,1)$, the indices
$\mu,\nu$ run from 1 to 4, and the mapping between tensor indices
and two-component spinor indices in defined by
\begin{equation}
V^\ada=\pmatrix{V^{1\dot1} & V^{1\dot2}\cr
                V^{2\dot1} & V^{2\dot2}\cr}
       ={1\over \sqrt2} \pmatrix{V^1 + V^4 & V^2 - V^3\cr
                                 V^2 + V^3 & -V^1 + V^4\cr} \ ,
\end{equation}
where $V^\mu$ is an arbitrary vector.  The van der Waerden symbols
$\sigma^\ada_\mu$ thus defined satisfy $\sigma^\ada_\mu\,
\sigma^{\mu\bdb} = \epsilon^{\alpha\beta}\, \epsilon^{\dot\alpha\dot\beta}$,
where $\epsilon_{12}=\epsilon^{12}=1$.

    The $N=2$ currents in this matter system are given by
\begin{eqnarray}
T &=& -\ft12 \del X^\ada\, \del X_\ada - \ft12 p_\a\, \del\t^\a +
                  \ft12 \del p_\a\, \t^\a \nonumber \\
G^{\dot1} &=& \t_\a\, \del X^{\a\dot1}\nonumber \\
G^{\dot2} &=& p_\a\, \del X^{\a\dot2} \label{n2diag}\\
J &=& p_\alpha\, \t^\alpha \ .\nonumber
\end{eqnarray}
Note that we are describing the $(2,2)$ spacetime in terms of a real
structure rather than the complex structure.  Here, one uses double
number of the form $x+e\, y$, where $e^2 =1$, rather than complex
numbers.   However, all the results we obtain in this real structure also
apply to the complex structure.

     Turning now to the BRST operator, we begin by introducing the
anticommuting ghosts $(b,c)$ and $(\beta,\gamma)$ for the bosonic currents
$T$ and $J$, and the commuting ghosts $(r^\da, s_\da)$ for the fermionic
currents $G^\da$, with $r^\da (z)\, s^{\dot\beta}(w) \sim
\epsilon^{\da\dot\beta} (z-w)^{-1}$.  As usual in a theory with fermionic
currents, it is necessary to bosonise the associated commuting ghosts. Thus
we write
\begin{equation}
r^{\dot\alpha} = \del\xi^\da\, e^{-\phi_\da} \ ,\qquad
s_\da =\eta_\da \, e^{\phi_\da}\ ,\label{fermionisation}
\end{equation}
where $\eta_\da$ and $\xi^\da$ are anticommuting fields with spin
1 and 0 respectively.  The OPEs of the bosonising fields are
$\eta_\da (z)\, \xi^{\dot\beta} (w) \sim \delta_\da^{\dot\beta}\,
(z-w)^{-1}$, and $\phi_\da(z)\, \phi_{\dot\beta} (w) \sim
  -\delta_{\da\dot\beta}\log (z-w) $.

    The BRST operator in terms of the bosonised fields is given by
\begin{eqnarray}
Q &=& c\Big( -\ft12 \del X_\ada\,\del X^\ada
             -\ft12 p_\a\,\del\t^\a +\ft12\del p_\a\,\t^\a
               \nonumber\\
&{}& -\ft12(\del\phi_1)^2 -\ft12 (\del\phi_2)^2 -\del^2\phi_1 -
\del^2\phi_2 -\beta\, \del\gamma - \eta_\da\,\del\xi^\da -b\,\del c
\Big)\nonumber \\
&{}& +\eta_1 e^{\phi_1} \t_\a\, \del X^{\a\dot1}
 - \eta_2 e^{\phi_2} p_\a\, \del X^{\a\dot2} + b\eta_1\eta_2\, e^{\phi_1 +
\phi_2}\label{diagbrst}\\
&{}& +\ft12\beta \Big(\eta_1\,\del\eta_2 -\del\eta_1\,\eta_2 -
\eta_1\eta_2(\del\phi_1 -\del\phi_2)\Big)\, e^{\phi_1 +\phi_2} +
\gamma(-p_\a\t^\a -\del\phi_1 + \del\phi_2)\ .\nonumber
\end{eqnarray}
(It is to be understood that an expression such as $e^{\phi_1 + \phi_2}$
really means $:e^{\phi_1}::e^{\phi_2}:$ with cocycle factors suppressed.  We
have also dropped the dots on the indices of the bosonised fields when they
do not appear in a covariant form, and suppressed the integration
symbol.)

     Since the zero modes of the $\xi^{\da}$ fields are not included in the
Hilbert space of physical states, there exist BRST non-trivial
picture-changing operators $Z^\da =\{Q, \xi^\da\}$ which can give new BRST
non-trivial physical operators when normal ordered with others.  Explicitly,
they take the form
\begin{eqnarray}
Z^1 &=& c\del \xi^1 - \t_\a\,\del X^{\a\dot1}\, e^{\phi_1} +
\Big(\beta\,\del\eta_2 + \ft12 \del\beta\, \eta_2 + \beta\,\eta_2\,
\del\phi_2 -b\,\eta_2 \Big)\, e^{\phi_1 +\phi_2}\ ,\label{diagz1}\\
Z^2 &=& c\del \xi^2 - p_\a\,\del X^{\a\dot2}\, e^{\phi_2} +
\Big(\beta\,\del\eta_1 + \ft12 \del\beta\, \eta_1 + \beta\,\eta_1\,
\del\phi_1 + b\,\eta_1 \Big)\, e^{\phi_1 +\phi_2}\ .\label{diagz2}
\end{eqnarray}
Unlike the picture-changing operator in the usual $N=1$ superstring, these
operators are not invertible.  This feature is not always true for the $N=2$
string; it depends on the choice of the basis for the fermionic currents. As
we shall see later, this has important consequences for the physical
spectrum.

\subsection{Physical Spectrum}

\indent     In this subsection, we shall only consider the physical spectrum
in the Neveu-Schwarz sector.  The simplest physical state is of standard
ghost structure, given by
\begin{equation}
V=c\, e^{-\phi_1 -\phi_2}\, e^{i p\cdot X}\ ,\label{vstate}
\end{equation}
which is annihilated by the BRST operator provided that the mass-shell
condition $p_\ada\,p^\ada=0$ is satisfied.   The conjugate operator
\begin{equation}
V^\dagger = \del c\, c\,\gamma\, e^{-\phi_1 -\phi_2}\,
e^{-i p\cdot X}\ ,\label{vdagger}
\end{equation}
is chosen by the requirement that the following inner product be
non-vanishing:
\begin{equation}
\Big\langle \del^2c\,\del c\, c\, \gamma\, e^{-2\phi_1 -2\phi_2}\Big\rangle
\ne 0\ .\label{innerproduct1}
\end{equation}
It is worth mentioning that the physical operator given in (\ref{vstate}) is
one member of a quartet, whose other three members correspond to replacing
the $c$ ghost by $\del c\,c$, $c\, \gamma$ or $\del c\,c\, \gamma$
respectively. Since all the physical operators that we shall consider in
this paper can occur with or without $\gamma$, we shall in general suppress
the $\gamma$ in the non-vanishing inner product.

    The physical operator given in (\ref{vstate}) is generally believed to
be the only physical degree of freedom of the theory in the NS sector. All
other BRST non-trivial NS operators that are annihilated by $Q$ are believed
to be related to $V$ by picture changing, and thus can be identified with
the same physical degree of freedom.   In other words, one can view two
physical operators $U$ and $V$ as equivalent if they satisfy
\begin{equation}
(Z^1)^{n_1}\, (Z^2)^{n_2}\, U = \lambda\, (Z^1)^{m_1}\,
(Z^2)^{m_2}\, V\label{identify}
\end{equation}
for non-singular scaling factor $\lambda$,  where $n_1, n_2, m_1, m_2$ are
integers.  The reason why one can make such an identification is that a
correlation function is independent of the locations of the picture-changing
operators \cite{fms}, and thus $U$ and $V$ will give rise to the same
correlation function.

     If the picture-changing operators are invertible, it follows that
$\lambda$ in (\ref{identify}) is non-singular since an invertible physical
operator cannot give rise to a trivial operator by normal-ordering with
another non-trivial physical operator \cite{w3cohomology}.  However, as
mentioned in the previous subsection, the picture-changing operators given
in (\ref{diagz1}) and (\ref{diagz2}) in this diagonal basis of fermionic
currents are not invertible.   The scaling factor $\lambda$ in
(\ref{identify}) is singular for certain physical operators and hence
they can not be identified as equivalent any more.  To
illustrate this, let us consider the following examples:
\begin{equation}
\Psi = h_\a\, c\, \t^\a\, e^{-\phi_2}\, e^{ip\cdot X}\ ,\qquad
\Phi = h_\a\, c\,  p^\a\, e^{-\phi_1}\, e^{ip\cdot X}\ .\label{twoexamp}
\end{equation}
They are physical operators provided that
\begin{equation}
p_\ada\, p^\ada =0\ ,\label{mass}
\end{equation}
and that the polarisation spinor $h_\a$ satisfies
\begin{equation}
p^\ada\, h_\a =0 \ .\label{polarcon}
\end{equation}
The mass-shell condition (\ref{mass}) implies that $p^{\a\dot1} =
\lambda\, p^{\a\dot2}$ for some constant $\lambda$.  It follows from
(\ref{polarcon}) that $h_\a$ only has one physical degree of freedom, which is
generically proportional to either $p^{\a\dot1}$ or $p^{\a\dot2}$.  In the
case when $p^{\a\dot1} = 0$ or $p^{\a\dot2} =0$, it follows that $\lambda$
becomes singular.  Under these circumstances, $h_\a$ can be only written as a
multiple of either $p^{\a\dot2}$ or $p^{\a\dot1}$ respectively.  When
$p^{\a\dot1} \ne 0$ or $p^{\a\dot2} \ne 0$, the physical operators $\Psi$
and $\Phi$ can be obtained by normal-ordering $Z^1$ or $Z^2$ with the
physical operator $V$ given by (\ref{vstate}):
\begin{equation}
Z^1\, V = p^{\a\dot1}\, c\, \t_\a\, e^{-\phi_2}\, e^{ip\cdot X}\ ,\qquad
Z^2\, V = - p^{\a\dot2}\, c\, p_\a\, e^{-\phi_1}\, e^{ip\cdot X}\ .
\end{equation}
However, when $p^{\a\dot1}=0$ or $p^{\a\dot2}=0$, the normal-ordered
products $Z^1\, V$ and $Z^2\, V$ vanish respectively. In either case, $\Psi$
or $\Phi$ describe an independent degree of freedom, which cannot be related
to $V$ by picture changing.  This corresponds to a case where $\lambda$
becomes singular in (\ref{identify}).

      The operators (\ref{twoexamp}) are two examples of an infinite number
of physical operators that are identified under picture changing for generic
on-shell momenta, but that become independent in special cases.  Consider
the following scalar-type operators:
\begin{eqnarray}
V_n &=& c\, (\del^{n-1}\t)^2\, (\del^{n-2}\t)^2\cdots \t^2\, e^{(n-1)\phi_1-
      (n+1)\phi_2}\, e^{ip\cdot X} \ ,\nonumber\\
U_n &=& c\, (\del^{n-1}p)^2\, (\del^{n-2}p)^2\cdots p^2\, e^{-(n+1)\phi_1+
      (n-1)\phi_2}\, e^{ip\cdot X} \ ,\label{scalar}
\end{eqnarray}
which are all physical provided that the mass-shell condition (\ref{mass})
is satisfied. Note that $p^2 =p_\a\, p^\a$, {\it etc}.
In addition, we introduce the spinor-type operators
\begin{eqnarray}
\Psi_n &=& h_\a\, c\, \del^{n-1}\t^\a \, (\del^{n-2}\t)^2\cdots \t^2\,
            e^{(n-1)\phi_1 - n\phi_2}\, e^{ip\cdot X} \ ,\nonumber\\
\Phi_n &=& h_\a\, c\, \del^{n-1} p^\a \, (\del^{n-2}p)^2\cdots p^2\,
            e^{-n\phi_1 + (n-1)\phi_2}\, e^{ip\cdot X} \ ,\nonumber\\
\widetilde\Psi_n &=& h_\a\, c\, \del^{n-1}\t^\a \,
                                 (\del^{n-2}\t)^2\cdots \t^2\,
            e^{(n-2)\phi_1 - (n+1)\phi_2}\, e^{ip\cdot X} \ ,\label{spinor}\\
\widetilde\Phi_n &=& h_\a\, c\, \del^{n-1} p^\a \,
                                   (\del^{n-2}p)^2\cdots p^2\,
            e^{-(n+1)\phi_1 + (n-2)\phi_2}\, e^{ip\cdot X} \ ,\nonumber
\end{eqnarray}
which are physical provided that the conditions (\ref{mass}) and
(\ref{polarcon}) are satisfied.  Note that the operators $V$, $\Psi$ and
$\Phi$ that we discussed previously are the special cases $V_0$, $\Psi_1$
and $\Phi_1$ respectively.

     For generic on-shell momentum, where neither $p^{\a\dot1}$ nor
$p^{\a\dot2}$ vanishes, all the above physical operators can be identified
as equivalent under picture changing.  Specifically, we have
\begin{eqnarray}
Z^1\, V_n &=& \Psi_{n+1} = Z^2\, V_{n+1} \ ,\nonumber\\
Z^1\, U_n &=& \Phi_n = Z^2\, U_{n-1} \ ,\nonumber\\
Z^1\,\widetilde\Psi_n &=& V_n= Z^2\,\widetilde\Psi_{n+1}\ ,\label{chain}\\
Z^1\,\widetilde\Phi_{n+1} &=& U_n = Z^2\, \widetilde\Phi_n\ ,\nonumber
\end{eqnarray}
where we have suppressed the non-singular scaling factors.  Note that if we
bosonise the $\t^\a$ and $p_\a$ fields, all the above physical operators are
pure exponentials.  We may then continue the indices to arbitrary negative
as well as positive integers.  In fact now we have the identities
$U_n=V_{-n}$, $\Phi_n=\Psi_{-n-1}$ and $\widetilde\Phi_n =
\widetilde\Psi_{-n-1}$.   This means that the equations (\ref{chain}) can be
reduced to just the first and third equations, giving a chain of
identifications for all these physical operators. This chain breaks down if
$p^{\a\dot1}=0$ or $p^{\a\dot2}=0$, since then $Z^1$ or $Z^2$ respectively
will annihilate all of the physical operators on which they act.  Thus there
are an infinite number of physical operators in the spectrum.    These
physical operators, although not equivalent for the special values of
momenta, can be related by picture changing for generic on-shell momenta.
One must therefore make a choice as to which is to be selected as the
representative that describes the massless spacetime field in the $N=2$
string theory. The requirement of this massless operator is that one can
construct all correlation functions with non-vanishing inner product by
multiple insertions of only this operator together with the necessary
picture-changing operators.  Then the NS spectrum of the $N=2$ string in the
diagonal basis is described by this massless operator that is well-defined
for all on-shell momenta, together with an additional infinite number
of massless operators that are defined for $p^{\a\dot1}=0$ or
$p^{\a\dot2}=0$.

\section{$N=2$ String in Off-diagonal Bases}

\indent  Our discussion in the previous section was for the case where the
fermionic currents of the $N=2$ algebra were the $\pm$ eigenstates under the
$U(1)$ current.   In this section, we investigate the $N=2$ string with an
off-diagonal basis of fermionic currents.   Contrary to naive expectations,
such a change of basis can actually affect the physical spectrum of the
theory.  We shall discuss this further at the end of the section.

     Since the fermionic currents appear in the BRST operator in the form
$s_\da\, G^\da$, the change of the basis of $G^\da$ is equivalent to
a change of basis for the ghosts $s_\da$.   For our purposes, the following
canonical transformation is sufficiently general:
\begin{eqnarray}
s_1 \longrightarrow s_1 +  x_1\, s_2\ ,&&\qquad
r^1 \longrightarrow {1\over 1- x_1\, x_2}
(r^1 -  x_2 \, r^2)\ ,\nonumber \\
s_2 \longrightarrow s_2 +  x_2 \, s_1\ ,&&\qquad
r^2 \longrightarrow {1\over 1- x_1\, x_2}
(r^2 -  x_1 \, r^1 )\ .\label{redefintion}
\end{eqnarray}
where $x_1, x_2$ are arbitrary constants.  After the bosonisation of the new
ghost fields by using (\ref{fermionisation}), the BRST operator for the
off-diagonal $N=2$ string becomes
\begin{eqnarray}
\widetilde Q &=& c\Big(-\ft12 \del X_\ada\, \del X^\ada -\ft12 p_\a\, \del
\t^\a + \ft12 \del p_\a\, \t^\a -\eta_\a\, \del\xi^\a -\beta\, \del\gamma
\nonumber\\
&-& \ft12 (\del\phi_1)^2 -\ft12 (\del\phi_2)^2 -\del^2\phi_1 -\del^2\phi_2
-b\,\del c \Big) + \eta_1\,\t_\a\,\del X^{\a\dot1}\, e^{\phi_1}
\nonumber\\
&-& \eta_2\, p_\a\, \del X^{\a\dot2}\, e^{\phi_2} -  x_1\, \eta_2\,
\t_\a\, \del X^{\a\dot1}\, e^{\phi_2} +
x_2\,\eta_1\, p_\a\, \del X^{\a\dot2}\, e^{\phi_1}\nonumber\\
&+&(1+x_1\, x_2) b\, \eta_1\, \eta_2\, e^{\phi_1 + \phi_2}
+ x_1\, b\, \del\eta_2\, \eta_2\, e^{2\phi_2}
+ x_2\, b\, \del\eta_1\, \eta_1 e^{2\phi_1}\label{offdiagbrst}\\
&+& \ft12 (1-x_1\, x_2)\beta\Big( \del\eta_1\, \eta_2 -\eta_1\,
\del\eta_2 + \eta_1\, \eta_2 (\del\phi_1-\del\phi_2)\Big)
e^{\phi_1 + \phi_2} \nonumber\\
&+&\gamma\Big(-p_\a\, \t^\a + {1+x_1\, x_2 \over 1- x_1\, x_2}
(\del\phi_1 -\del\phi_2) \nonumber\\
&-& {2 x_1 \over 1-x_1\, x_2} \eta_2\, \del\xi^1
\, e^{-\phi_1 + \phi_2} - {2 x_2 \over 1 -x_1\, x_2}\eta_1\, \del\xi^2\,
e^{\phi_1 -\phi_2} \Big)\ ,\nonumber
\end{eqnarray}
The picture-changing operators in this generic basis
are given by
\begin{eqnarray}
\widetilde Z^1 &=& c\,\del\xi^1 + \ft12(1- x_1\, x_2)
\Big(\del\beta\,\eta_2 + 2\beta\,\del\eta_2 + 2 \beta\,
\eta_2\,\del\phi_2 \Big) e^{\phi_1 + \phi_2}\nonumber\\
&+& x_2\Big(\del b\,\eta_1 + 2b\,\del\eta_1 + 2 b\,\eta_1\,\del\phi_1
\Big) e^{2\phi_1} - (1+ x_1\, x_2) b\,\eta_2\,
e^{\phi_1+\phi_2}\label{tildez1}\\
&+&{2 x_2\over 1- x_1\, x_2}\, \gamma\,\del \xi^2
\,e^{\phi_1 - \phi_2} -  x_2\, p_\a\,\del X^{\a\dot2}\, e^{\phi_1}
-\t_\a\,\del X^{\a\dot1}\,e^{\phi_1}\ ,\nonumber\\
\widetilde Z^2 &=& c\,\del\xi^2 + \ft12(1- x_1\, x_2)
\Big(\del\beta\,\eta_1 + 2\beta\, \del\eta_1 + 2 \beta\,
\eta_1\,\del\phi_1\Big) e^{\phi_1 + \phi_2}\nonumber\\
&+& x_1\Big(\del b\,\eta_2 + 2b\,\del\eta_2 + 2 b\,\eta_2\,\del\phi_2
\Big) e^{2\phi_2} + (1+ x_1\, x_2) b\,\eta_1\,
e^{\phi_1+\phi_2}\label{tildez2}\\
&+&{2 x_1\over 1- x_1\, x_2}\, \gamma\,\del \xi^1
\,e^{-\phi_1 + \phi_2} -  x_1\, \t_\a\,\del X^{\a\dot1}\, e^{\phi_2}
-p_\a\,\del X^{\a\dot2}\,e^{\phi_2}\ .\nonumber
\end{eqnarray}
The terms involving $e^{2\phi_1}$ and $e^{2\phi_2}$ in (\ref{tildez1}) and
(\ref{tildez2}), which are absent in the diagonal basis, $x_1=x_2=0$, are
crucial for the invertibility of these operators. If $ x_2\ne 0$ or $x_1 \ne
0$, then $(\widetilde Z^1)^{-1}$ or $(\widetilde Z^2)^{-1}$ exist, given by
\begin{equation}
(\widetilde Z^1)^{-1} ={1\over  x_2}\, c\, \del\xi^1\, e^{-2\phi_1} +
{\rm more}\ , \qquad
(\widetilde Z^2)^{-1} ={1\over  x_1}\,c\, \del \xi^2\,
e^{-2\phi_2} + {\rm more} \label{picleading}
\end{equation}
respectively.   As discussed in Ref.~\cite{BKL} for the special case where $
x_1=- x_2 =1$, the leading terms of the inverse picture-changing operators
are not by themselves BRST invariant, but one can add an infinite number of
additional pure ghost terms to achieve BRST invariance.   We find that this
is true also for arbitrary non-singular values of $x_1$ and $ x_2$, although
the structure of the inverse operators is more complicated.   Details can be
found in the Appendix.

     We now turn to the spectrum of the $N=2$ string in this off-diagonal
basis.  The operator $V$ given in (\ref{vstate}) is still physical provided
the mass-shell condition (\ref{mass}) is satisfied.  The operators that can
be obtained from the chains of identification (\ref{chain}) no longer have
the simple forms given in (\ref{scalar}) and (\ref{spinor}).   For example,
we now have,
\begin{eqnarray}
\widetilde Z^1 V &=& \Big(p^{\a\dot1}\,c\,\t_\a\, e^{-\phi_2} +
 x_2\, p^{\a\dot2}\, c\, p_\a\, e^{-\phi_2} - x_2\,
\eta_1\, e^{\phi_1 - \phi_2} -{2 x_2\over 1 -  x_1 x_2}
c\, \gamma\, \del \xi^2 \, e^{-2\phi_2}\Big) e^{ip\cdot X}\ ,\label{tzv}\\
\widetilde Z^2 V &=&- \Big(p^{\a\dot2}\,c\,p_\a\, e^{-\phi_1} +
 x_1\, p^{\a\dot1}\, c\, \t_\a\, e^{-\phi_1} + x_1\,
\eta_2\, e^{-\phi_1 + \phi_2} +{2 x_1\over 1 -  x_1 x_2}
c\, \gamma\, \del \xi^1 \, e^{-2\phi_1}\Big) e^{ip\cdot X}\ ,\nonumber
\end{eqnarray}
One can clearly see the differences between the $\widetilde Z^1$, $\widetilde
Z^2$ operators and the $Z^1$, $Z^2$ operators which are the special cases
for the diagonal basis $ x_1= x_2=0$.   The $Z^1$ or $Z^2$ operators
annihilate physical operators when $p^{\a\dot1} = 0$ or $p^{\a\dot2}=0$
respectively.  On the other hand, we see that $\widetilde Z^1 V$ and $
\widetilde Z^2 V$ are non-zero for all on-shell momenta, as must be the case
since $\widetilde Z^1$ and $\widetilde Z^2$ have inverses for non-singular
values of $ x_2$ and $ x_1$.

    The discussion of the cohomology becomes delicate when one has physical
operators that involve infinite sums of terms, which occurs in the
off-diagonal bases.  For example, this is the case for the conjugates of the
operators (\ref{tzv}), for all on-shell momenta.  On the other hand, the
operator $\widetilde Z^1 V$ can be written as the BRST commutator
$\{\widetilde Q, \chi\}$ where $\chi$ itself involves an infinite sum of
terms, for the special momentum $p^{\a\dot1}=0$.  Thus one has to define
carefully the class of operators that one allows in the Hilbert space and
its conjugate space in order to discuss the the BRST triviality of a
physical state.  The most natural choice is to require that the Hilbert
space of states involve operators with only finitely many terms, while the
space of the conjugate states can involve operators with infinitely many
terms. This avoids the occurrence of infinite sums in the evaluation of
inner products.  Thus a physical operator $V$ with finitely many terms is
only BRST trivial if it is expressible as $\{\widetilde Q,S\}$ for some $S$
that also has a finite number of terms.   Note that the inverse picture
changing operators are allowable, even though they have infinitely many
terms, since they live in the space of conjugate operators.   With this
definition of allowable states, the cohomology in the off-diagonal bases
differs from the cohomology in the diagonal basis.   One way to understand
this difference is to note that a physical operator with finitely many terms
in the diagonal basis may map into a physical operator with infinitely many
terms in the off-diagonal bases, and vice versa. In the off-diagonal bases,
consistency requires us to allow physical states in the conjugate space that
have infinitely many terms; however, in the diagonal basis, it is consistent
to consider physical states with only finitely many terms in the whole
Hilbert space of states and conjugates.

    To summarise, for the $N=2$ string in a completely off-diagonal basis,
{\it i.e.~}where $ x_1\, x_2\ne 0$, both picture-changing operators are
invertible, and thus the physical spectrum in the NS sector is described by
a single massless state.  There is also a massless R operator in the
spectrum which we shall discuss in section 5.  In the diagonal basis, where
$ x_1= x_2 =0$, neither picture-changing operator is invertible, and there
are additional physical states in the NS spectrum with further momentum
constraints. These were discussed in detail in section 2.   There are
intermediate semi-diagonal cases, where either $ x_1=0$ or $ x_2=0$.  In
these cases, one picture-changing operator is invertible whilst the other is
not.   Under these circumstances, half of the additional physical states
discussed in section 2 can be identified under the invertible
picture-changing operator.

\section{Interactions in the Diagonal Basis}

\subsection{Inner Product}

\indent The physical spectrum of a theory is defined in the BRST formalism
by the set of operators that are closed but not exact under the BRST
operator.  One needs to define an inner product to establish a pairing of
physical states and their conjugates.  The choice of an inner product
for a consistent theory is not in general unique.  A necessary requirement
is that no BRST trivial operator should have a non-zero inner product
with any physical operator and that every BRST non-trivial physical operator
should have a conjugate pair with which it has non-vanishing norm.  In a
string theory, there is a simple procedure for defining an inner-product
that meets this requirement.   It can be summarised as the requirement
of momentum conservation in the functional integral over the products of
operators.    For the spacetime coordinates, it is just the usual
conservation law of spacetime momentum.  For the fermionic matter and the
ghost sector, one can bosonise these fields and again require the
conservation of momentum, which is usually modified by the background
charges of the bosonising fields. (This requirement should not be imposed
for fields such as $\xi^{\dot\a}$, whose zero modes are excluded from the
Hilbert space.)

     The inner product given in (\ref{innerproduct1}) is the one for the
$N=2$ string that follows by applying the above procedure.  To be more
precise, it is the one that corresponds to taking the $N=2$ algebra in its
standard untwisted form, where both fermionic currents have spin $\ft32$.
However, the $N=2$ algebra can be twisted by redefining the
energy-momentum tensor
\begin{equation}
\widetilde T = T + \ft12(1 - s) \del J\ ,\label{newt}
\end{equation}
for arbitrary $s$.  In the diagonal basis, but not in a generic basis, the
fermionic currents remain primary under this redefined energy-momentum
tensor, albeit with different conformal spins $(2-\ft12 s)$ and $(1+\ft12
s)$.  In our realisation (\ref{n2diag}), this corresponds to $(p_\a, \t^\a)$
having spins $(\ft12 s, 1-\ft12 s)$.  The spins of the ghost fields $(c, b)$,
$(\gamma, \beta)$ and $(\eta_\da, \xi^\da)$ are unchanged,  and the
energy-momentum tensor for the $\phi_1$ and $\phi_2$ fields becomes $T_\phi
= -\ft12 (\del\phi_1)^2 -\ft12 (\del \phi_2)^2 + \ft12(s-3)\del^2\phi_1
-\ft12(s+1)\del^2\phi_2$.  If we bosonise the $(p_\a, \t^\a)$ fields as
\begin{equation}
p_\a = e^{-i\sigma_\a}\ ,\qquad \t^\a = e^{i\sigma_\a}\ ,\label{bosept}
\end{equation}
then the requirement of conservation of momentum in the functional integral
leads to the following inner product:
\begin{equation}
\Big\langle \del^2 c\,\del c\, c\,\gamma\,
e^{i(s-1)\sigma_1 + i(s-1)\sigma_2 + (s-3)\phi_1 -
(s+1)\phi_2}\Big\rangle \ne 0\ .\label{innerproduct2}
\end{equation}
If $s$ lies in the range $-2\le s\le 4$, all the matter currents have
non-negative spin.   For interactions involving only the Neveu-Schwarz
sector, $s$ has to take integer values, in which case, the inner product
defined in (\ref{innerproduct2}) can be re-expressed in terms of $p_\a$ or
$\t^\a$.

      In the BRST language, this twisting of the algebra corresponds to a
canonical field redefinition of the ghost fields $(c, b)$ and $(\gamma,
\beta)$, whose spins are independent of the twisting:
\begin{eqnarray}
c\longrightarrow c \ ,&&\qquad \gamma \longrightarrow \gamma +
\ft12(1-s)\del c\ ,\nonumber\\
\beta\longrightarrow \beta\ ,&&\qquad b \longrightarrow b +
  \ft12(1-s)\del\beta\ .
\end{eqnarray}
Thus we can reinterpret (\ref{innerproduct2}) as defining different
choices of inner product for the same untwisted $N=2$ string.  One can
easily verify that all the physical operators given by (\ref{scalar}) and
(\ref{spinor}) have their conjugate pairs under this general definition
of the inner product (\ref{innerproduct2}).

     In the above discussion of inner products, we considered the theory in
a gauge-fixed background.  Eventually, in the construction of the string
theory, one integrates over the moduli for the world-sheet gauge fields.  In
particular, the $U(1)$ moduli for the spin--1 current $J^{\rm tot}\equiv
\{Q,\beta\}$ are present even at the tree level, since they correspond to
integrations over instanton sectors associated with the points where the
external states puncture the sphere.  Since there is a term $\int d^2z A
J^{\rm tot}$ in the world-sheet Lagrangian, integration over the moduli for
the $U(1)$ gauge field $A$ reduces to the insertion of an operator
\begin{equation}
N_\lambda\equiv e^{\lambda \int^z J^{\rm tot}} = (1+\lambda\, c\beta)
e^{\lambda(+i \sigma_1 +i \sigma_2 -\phi_1 +\phi_2)}\label{mod}
\end{equation}
at each vertex, where $\lambda$ is a constant parameter which is eventually
integrated over.  In fact the physical operators $V_n$ and $U_n$ given in
(\ref{scalar}), which in bosonised form are given by
\begin{equation}
Y_t=c\, e^{i(t-1)\sigma_1 + i(t-1)\sigma_2 +(t-2)\phi_1 -t \phi_2}\, e^{ip
\cdot X}, \label{bosonisedUV}
\end{equation}
with $V_n= Y_{n+1}$ and $U_n=Y_{-n+1}$, are all related by the operator
$N_\lambda$ defined in (\ref{mod}), since we can, for example, write
$Y_t=(N_{t-1}\, Y_1)$.    The parametrisation of different inner products that
we described above corresponds precisely to the parametrisation of the
$U(1)$ moduli.  Thus we may, without loss of generality, always choose the
$s=1$ inner product (\ref{innerproduct1}), and from now on, we shall assume
that this choice is made.

     It should be emphasised that the above discussion applies only to the
diagonal basis for the fermionic currents.  In the off-diagonal bases,
$J^{\rm tot}$ is no longer expressible as the derivative of a local
expression built from the bosonising fields, and so no local form for the
operator $N_\lambda$ in (\ref{mod}) can be given.  Correspondingly, there
is no freedom to choose different inner products in an off-diagonal basis.

\subsection{Amplitudes}

\indent In the BRST formalism, there is a necessary condition for building a
non-vanishing $n$-point function, {\it i.e.~}the conservation of momentum in
the functional integral.   This requires the conservation law of spacetime
momentum,  and that the product of the physical operators has to include the
structure that gives a non-vanishing inner product. If this condition is
satisfied, it becomes a matter of detailed computation to determine the
result.

      In sections 2 and 3, we obtained the physical spectrum in the NS sector
for the $N=2$ string, and showed that it is different in different bases for
the fermionic currents.   The physical spectrum for the $N=2$ string in
completely off-diagonal bases in the NS sector is described by a single
massless operator. When the fermionic currents are in the diagonal basis,
physical operators in different pictures cannot be identified owing to the
non-invertibility of the picture-changing operators.   In the subsection
we construct tree-level amplitudes for the $N=2$ string in the diagonal
basis.  We shall discuss the amplitudes for the off-diagonal bases in
section 5.

     The massless operators in the diagonal basis can be described by the
bosonised expression $Y_t$ given in (\ref{bosonisedUV}).  Different values
for the parameter $t$ correspond to different values for the $U(1)$ moduli.
In a tree-level string scattering amplitude, the values of the $t$
parameters will be integrated out.  If, for some choice of the parameters,
momentum conservation can be satisfied, then a non-vanishing amplitude can
result when the integrations are performed. To build a non-vanishing
three-point function, it is necessary to insert picture-changing operators
in order to be able to achieve momentum conservation.  In fact there are in
total three possibilities, namely the insertion of $Z^1\, Z^2$ or $Z^1\,
Z^1$ or $Z^2\, Z^2$.  For the first of these, we find that the three-point
amplitude takes the form
\begin{eqnarray}
&&\Big\langle Z^1 Y_1(z_1)\, Z^2 Y_1(z_2)\, Y_1(z_3)\Big\rangle = \nonumber\\
&&= \Big\langle p_{(1)}^{\a\dot1} c\, \t_\a e^{-\phi_2} e^{ip_{(1)}\cdot X}
(z_1)\,\,\, p_{(2)}^{\beta\dot2} c p^{\phantom{\chi}}_\beta
e^{-\phi_1} e^{ip_{(2)}\cdot X}(z_2)
\,\,\, ce^{-\phi_1-\phi_2} e^{ip_{(3)}\cdot X}(z_3)\Big\rangle
\label{vvv3pf}\\
&&=c_{12}\ ,\nonumber
\end{eqnarray}
where
\begin{equation}
c_{ij} = p_{(i)\,\alpha}{}^{\dot1}\,
         p_{(j)}{}^{\alpha\dot2} - p_{(j)\,\alpha}{}^{\dot1}\,
         p_{(i)}{}^{\alpha\dot2}\ .\label{cijs}
\end{equation}
The mass-shell condition (\ref{mass}) was used in order to write $c_{ij}$
in the antisymmetric form (\ref{cijs}).  We have made a specific choice for
the $t$ parameters for the $Y_t$ operators that gives the non-zero result.
In fact, the result is the same for any choice of the three parameters
$t_1$, $t_2$ and $t_3$ such that $t_1+t_2+t_3=3$.  For any other choice, the
result is zero.

    If instead we insert two $Z^1$ picture-changing operators in a
three-point function, then the momentum balance is modified and now the
three physical operators $Y_t$ must have parameters satisfying
$t_1+t_2+t_3=4$ in order to obtain a non-zero result.  The amplitude is then
given by
\begin{equation}
\Big\langle Z^1 Y_{t_1}(z_1)\, Z^1 Y_{t_2}(z_2)\, Y_{t_3}(z_3)\Big\rangle =
b_{12}\ ,\label{bij3pf}
\end{equation}
where
\begin{equation}
b_{ij} = p^{\phantom{X}}_{(i)\,\a}{}^{\dot1}\,
          p^{\phantom{X}}_{(j)}{}^{\a\dot1}\ .\label{bij}
\end{equation}
It is easy to see that the $b_{ij}$'s are antisymmetric.

    Finally, if two  $Z^2$ picture-changing operators are used, we obtain a
non-vanishing three-point amplitude for physical operators $Y_t$ whose
parameters satisfy $t_1+t_2+t_3=2$:
\begin{equation}
\Big\langle Z^2 Y_{t_1}(z_1)\, Z^2 Y_{t_2}(z_2)\, Y_{t_3}(z_3)\Big\rangle =
\tilde b_{12}\ ,\label{tbij3pf}
\end{equation}
where
\begin{equation}
\tilde b_{ij} = p^{\phantom{X}}_{(i)\,\a}{}^{\dot2}\,
          p^{\phantom{X}}_{(j)}{}^{\a\dot2}\ .\label{tbij}
\end{equation}

     The above three-point amplitudes, obtained by inserting the three
possible combinations of two picture-changing operators, represent the only
possible ways of achieving momentum conservation and hence a non-vanishing
result.  We observe that they give three different results, namely $c_{12}$,
$b_{12}$ and $\tilde b_{12}$.  Indeed, one can be zero while another is not.
Since the three-point amplitudes are different, it follows that physical
operators in different pictures cannot be identified.  This is consistent
with our discussion in section 2, where we saw that in the diagonal basis
the picture-changing operators are not invertible, and that consequently the
identification of physical operators in different pictures breaks down at
special values of the spacetime momenta.

   In the diagonal basis, the operator $Y_t$ is physical for all values of
$t$.  In particular, when $t$ is an integer it can be viewed as a
Neveu-Schwarz operator.  If instead we allow $t$ to take half-integer
values, it becomes a Ramond operator.  Since all the $Y_t$ operators are
related by the operator $N_\lambda$ given in (\ref{mod}), it suffices to
consider just the NS sector.  However, this does not imply that there is
only one physical operator in the spectrum, since, as revealed by the
three-point amplitudes, these physical operators give rise to different
interactions.   The physical operators with the same pictures as $Z^1 Y_t$
and $Z^2 Y_t$ cannot be identified with $Y_t$ itself, nor with each other
since there exist physical operators in these pictures even when $Z^1 Y_t$
and $Z^2 Y_t$ are zero at special momenta.  Determining the minimal set of
inequivalent physical operators is not easy in the diagonal basis, owing to
the pathology associated with the non-invertibility of the picture-changing
operators.  We shall postpone further discussion of this point until the
next section, where we shall consider the interactions in non-diagonal
bases, which do not suffer from this pathology.

     Let us now consider the four-point amplitudes in the diagonal basis.
Again, there are only three possible sets of picture-changing operators that
can be inserted in order to satisfy momentum conservation.  The three
possible amplitudes are given by
\begin{eqnarray}
&&\Big\langle Z^1Y\,\, Z^1Y\,\, \oint b\, Z^2 Y\,\,
                                Z^2 Y\Big\rangle =
(-stu + u\, c_{14}\, c_{23} + t\, c_{13}\, c_{24})
{\Gamma(-\ft12 s) \Gamma(-\ft12 t)\over \Gamma(\ft12 u)}\ ,
\nonumber\\
&&\Big\langle Z^1 Y\,\, Z^1 Y\,\, \oint b\, Z^1 Y\,\, Z^2 Y
\Big\rangle
=(-b_{14}\, us + b_{13}\, c_{24}\, s + b_{12}\, c_{34}\, u)
{\Gamma(-\ft12 s) \Gamma(-\ft12 t) \over \Gamma(\ft12 u)}\ ,
\label{bij4pf}\\
&&\Big\langle Z^2 Y\,\, Z^2 Y\,\, \oint b\, Z^2 Y\,\, Z^1 Y
\Big\rangle
=(\tilde b_{14}\, us + \tilde b_{13}\, c_{24}\, s +
\tilde b_{12}\, c_{34}\, u)
{\Gamma(-\ft12 s) \Gamma(-\ft12 t) \over \Gamma(\ft12 u)}\ ,
\nonumber
\end{eqnarray}
where $s$, $t$ and $u$ are the Mandelstam variables, {\it i.e.}
$p_{(i)\,\alpha}{}^{\dot1}\, p_{(j)}{}^{\alpha\dot2} +
p_{(j)\,\alpha}{}^{\dot1}\, p_{(i)}{}^{\alpha\dot2}$ for $(i,j) = (1,2)$,
$(1,4)$ and $(1,3)$ respectively.  In each case the $t$ parameters on the
physical operators $Y_t$ have been suppressed, and should be chosen so as to
achieve momentum conservation.

     These four-point amplitudes in fact vanish identically for kinematical
reasons, since the momentum-dependent prefactors are all zero:
\begin{eqnarray}
stu - u\, c_{14}\, c_{23} -t\, c_{13}\, c_{24} &=&0\ ,\nonumber\\
b_{14}\, us - b_{13}\, c_{24}\, s - b_{12}\, c_{34}\, u &=& 0\ ,
\label{bidentity}\\
\tilde b_{14}\, us + \tilde b_{13}\, c_{24}\, s +
\tilde b_{12}\, c_{34}\, u &=& 0\ .\nonumber
\end{eqnarray}
These identities are in
fact elementary consequences of the mass-shell condition (\ref{mass}) and
the conservation law of spacetime momentum in the four-point function:
The mass-shell condition implies that
\begin{equation}
p_{(i)\, \a}{}^{\dot2} = \lambda_{(i)}\, p_{(i)\, \a}{}^{\dot1}
\qquad {\rm for}\qquad i=1,2,3,4 \ .\label{masssol}
\end{equation}
where the $\lambda_{(i)}$'s are arbitrary constants.  Thus the conservation
law of momentum can be expressed as
\begin{equation}
\sum_{i=1}^4 p_{(i)\,\a}{}^{\dot1} = 0\ ,\qquad
\sum_{i=1}^4 \lambda_{(i)}\, p_{(i)\, \a}{}^{\dot1} = 0\ .
\label{momcon}
\end{equation}
These four equations can be used to solve four parameters.  Substituting
equation (\ref{masssol}) and the solved parameters into the right-hand sides
of the equations (\ref{bidentity}), one can easily show that they vanish
identically.  Note that we presented the proof for the case of a real
structure on the $(2,2)$ spacetime; the proof in the case of the complex
structure is analogous, giving a simpler derivation of the result obtained
in \cite{OV}.   Thus the vanishing of the four-point amplitudes in the $N=2$
string is a direct consequence of the $(2,2)$ signature, independent of the
choice of real or complex structure.

\section{Interactions in Off-diagonal Bases}

     In section 3 we described the $N=2$ string in a family of off-diagonal
bases for the spin--$\ft32$ currents, and obtained the physical spectrum in
the NS sector.  In these cases the picture-changing operators are
invertible, and hence the physical operators in different pictures can be
identified.  Thus there is only one physical degree of freedom in the NS
sector.   There is also a physical operator in the Ramond sector.  We find
that it is given by
\begin{equation}
W = c\,e^{-\ft12\phi_1 -\ft12\phi_2} \Big(
h_1\, i e^{\ft{i}2 \sigma_1 - \ft{i}2 \sigma_2} +
h_2\, e^{-\ft{i}2 \sigma_1 + \ft{i}2 \sigma_2}\Big)
e^{ip\cdot X} \equiv h_\a\, W^\a\ .\label{wstate}
\end{equation}
It is physical if the mass-shell condition (\ref{mass}) and the
spinor polarisation condition $p^\ada h_\a=0$ are satisfied.   Note that
this operator is physical in all the bases of the fermionic currents that
we discussed in sections 2 and 3, including the diagonal basis.  On the
other hand, the $Y_t$ operators given by (\ref{bosonisedUV}) are no longer
physical, except at $t=1$, in any of the off-diagonal bases.

     The Ramond operators in different pictures in the off-diagonal bases can
be identified by means of the picture-changing operators.  Thus there are
two physical operators in the spectrum.   They can be represented by two
operators whose forms are independent of the choice of basis for the
fermionic currents, namely the NS operator $V=Y_1$ given by (\ref{vstate}),
and the R operator $W$ given by (\ref{wstate}).

     In the diagonal basis, as we discussed in section 3, physical states in
the NS and R sectors can be related by the operator $N_\lambda$ given in
(\ref{mod}) that parametrises the $U(1)$ gauge-field moduli.  However such a
relation cannot be made in the off-diagonal bases, since then the operator
$N_\lambda$ is non-local in the bosonised field variables.  To see
this, we note from (\ref{offdiagbrst}) that the total $U(1)$ current is
given by
\begin{eqnarray}
J^{\rm tot} &=& \partial\Big(c\beta +i\sigma_1 + i\sigma_2 + {1+ x_1\,
x_2\over 1-x_1 \, x_2}(\phi_1-\phi_2) \Big)\nonumber\\
&&-{2 x_1\over 1- x_1\, x_2}\, \eta_2\, \del\xi^1\, e^{-\phi_1+\phi_2} -
{2 x_2\over 1-x_1\, x_2}\, \eta_1\, \del\xi^2\, e^{\phi_1-\phi_2} \ .
\label{jtot}
\end{eqnarray}
Since $N_\lambda$ is defined as $\exp(\lambda\int^z J^{\rm tot})$, we see
that although the terms in the first line in $J^{\rm tot}$ will integrate to
give a local expression, those in the second line, which are present in any
of the off-diagonal basis choices, will not.

     The upshot of the above discussion is that there appears to be no way
of identifying the NS and R physical operators in the off-diagonal bases.
This point is reinforced by looking at the pattern of interactions.  For the
three-point amplitudes, there are two possibilities, namely the interaction
of three NS operators, or the interaction of two R operators with one NS
operator.  For the first of these we have
\begin{equation}
\Big\langle Z^1 V\, Z^2 V\, V \Big\rangle = c_{12}\ ,\label{NSNSNS}
\end{equation}
just as in the diagonal basis, where $c_{12}$ is given by (\ref{cijs}).  For
the second of the three-point amplitudes, we find
\begin{equation}
\Big\langle W\, W\, V\Big\rangle = h_{(1)\alpha}\, h_{(2)}^\alpha\ .
\label{RRNS}
\end{equation}
Note that in this latter case no picture-changing operators are required,
and hence there is no explicit momentum dependence in the three-point
amplitude. Thus comparing the two three-point amplitudes, we see that they
are manifestly inequivalent.  In particular, the amplitude (\ref{NSNSNS})
vanishes for special values of the momenta, whereas the amplitude
(\ref{RRNS}) does not.

     The above remarks deserve further comment.  The polarisation spinor
$h_\alpha$ in the Ramond operator (\ref{wstate}) satisfies the
physical-state condition $p^{\alpha\dot\alpha}\, h_\alpha=0$.  For generic
on-shell momenta, the solution to this condition can be written as either
$h^\alpha\propto p^{\alpha\dot 1}$ or $h^\alpha\propto p^{\alpha\dot 2}$,
since these momentum components are proportional to each other on shell.
However, there is no way of explicitly solving the physical-state condition
for $h_\alpha$ in a way that remains non-singular for all allowable on-shell
momenta, including the cases where $p^{\alpha\dot1}=0$ or
$p^{\alpha\dot2}=0$.  Thus we see that the Ramond operator (\ref{wstate}) is
intrinsically spinorial in character, in the sense that it should properly
be described with a polarisation spinor $h_\alpha$ rather than as a scalar
like the NS operator (\ref{vstate}).

      The differences between the Ramond operator and the Neveu-Schwarz
operator are highlighted when one chooses, as we do in this paper, a {\it
real} structure on the $(2,2)$ spacetime, as opposed to the complex
structure used in \cite{OV}.  In the complex case, if one uses the
two-component spinor notation, the analogous momentum components
$p^{\alpha\dot1}$ and $p^{\alpha\dot2}$ are related by complex conjugation,
and thus the vanishing of one implies the vanishing of the other.  In that
case, therefore, one can identify the NS and R operators without
encountering inconsistencies, since their correlation functions can be
viewed as equivalent when the momenta are on shell.  However, the fact that
the NS and R operators cannot be identified in the case of the real
structure shows that there is no symmetry that forces one to make the
identification.  Thus the identification that is customarily made in the
case of the complex structure should be viewed as a consistent choice,
rather than as a requirement imposed by symmetry.

     The description of the Ramond sector in terms of the spinor operator
(\ref{wstate}), and its interaction with the NS operator, is also valid in
the diagonal basis for the fermionic currents.  Moreover, if we solve the
physical-state condition on the polarisation spinor by writing $h_\alpha$ in
terms of the on-shell momentum, the three-point function (\ref{RRNS}) becomes
a linear combination of $c_{12}$, $b_{12}$ and $\tilde b_{12}$, precisely
reproducing the interactions for the diagonal basis that we described in
section 4.  Thus the best way to organise the physical states in the diagonal
basis, which did not emerge in the presentation in terms of scalar-type
operators in sections 3 and 4, is by means of a scalar NS operator and a
spinorial R operator.  In the diagonal basis, the physical operators in the
NS and R sectors are related by the $U(1)$ modulus.  One might
therefore think that the physical spectrum in the theory is described by only
one massless operator.  However, owing to the non-invertibility of the
picture-changing operators, the physical operators at the different pictures
cannot be identified for all on-shell momenta.  The three-point interactions
of all the physical operators yield precisely the three types of amplitudes
$c_{12}$, $b_{12}$ and $\tilde b_{12}$.  Thus the physical degrees of
freedom in the diagonal basis can also be effectively described by a scalar
NS operator and a spinorial R operator.

    There is also a further advantage that results from this way of
describing the physical operators in the diagonal basis:  Although the
manifest $SO(2,2)=SL(2,R)_{\rm L}\times SL(2,R)_{\rm R}$ Lorentz covariance
is broken by the $N=2$ constraints, since the $SL(2,R)_{\rm L}$ covariance of
the dotted indices is broken, the momentum-dependent quantity $c_{12}$
arising in the three-point amplitude for the NS operators does at least
preserve an $SO(1,1)$ subgroup of $SL(2,R)_{\rm L}$.  On the other hand,
this subgroup is broken by $b_{12}$ and $\tilde b_{12}$.  However, if we
instead describe the additional physical degrees of freedom by the
spinorial-type R operator (\ref{wstate}), the breaking of the residual
$SO(1,1)$ is avoided.

     So far we have focussed our attention on the holomorphic sector of the
$N=2$ string.  For a theory of closed strings, this should be combined with
the corresponding anti-holomorphic sector.  This leads to four physical
operators, namely
\begin{equation}
\Phi = V\, \overline V\ ,\qquad F=h_{\a\beta}\, W^\a\, \overline W^\beta\ ,
\qquad \chi = h_\a\,  V\, \overline W^\a\ ,\qquad
\psi = g_\a\, W^\a\, \overline V\ ,\label{off4fields}
\end{equation}
where $W^\a$ is defined in (\ref{wstate}), $h_{\a\beta}=h_{\beta\a}$, and
the barred operators are in the anti-holomorphic sector.   The operators
(\ref{off4fields}) are physical provided that the mass-shell condition
(\ref{mass}) is satisfied and that $p^\ada\, h_\a =0$, $p^\ada\, g_\a =0$
and $p^\ada\, h_{\a\beta} =0$.   The three-point amplitudes that can be
built from these operators are
\begin{eqnarray}
&&\Big\langle \Phi\, \Phi\, \Phi\Big\rangle = c_{12}^2\ ,\qquad
\Big\langle \chi\, \chi\, \Phi\Big\rangle = c_{12}\,
 h_{(1)\a}\, h^\a_{(2)}\ ,\qquad
\Big\langle \psi\, \psi\, \Phi\Big\rangle =c_{12}\,
 g_{(1)\a}\, g^\a_{(2)}\ ,\nonumber\\
&&\Big\langle F\, F\, \Phi \Big\rangle = h_{(1)\a\beta}\, h_{(2)}^{\a\beta}
\ ,\qquad \Big\langle F\, \chi\, \psi\Big\rangle =
h_{(1)\a\beta}\, h_{(2)}^\a\, g_{(3)}^\beta\ .\label{off3pfs}
\end{eqnarray}
The Lagrangian for the spacetime fields $\{\Phi,
\chi_\a, \psi_\a, F_{\a\beta}\}$ associated with these operators is given
by
\begin{eqnarray}
{\cal L} &=& -\ft12\del_\ada\Phi\, \del^\ada\Phi + \zeta^{\dot\a}\,
\del_\ada \chi^\a + \lambda^{\dot\a}\, \del_\ada\psi^\a
+ \chi_\a\, \psi^\a + \Lambda^{\dot\a\dot\beta}\, \del_\ada
\del_{\beta\dot\beta} F^{\a\beta} -\ft12 F_{\a\beta} F^{\a\beta}\nonumber\\
&+& \ft13 \Phi\,\, \del_\a\del_{\tilde\beta} \Phi\,\,
\del^\a\del^{\tilde\beta} \Phi + \ft12 \Phi\,\,
\del_\a\del_\beta{}^{\dot\beta}\lambda_{\dot\beta}\,\, \del^{\tilde\alpha}
\del^{\beta\dot\gamma} \lambda_{\dot\gamma} + \ft12 \Phi\,\,
\del_\a\del_\beta{}^{\dot\beta}\zeta_{\dot\beta}\,\, \del^{\tilde\alpha}
\del^{\beta\dot\gamma} \zeta_{\dot\gamma} \label{offfield}\\
&+& \ft12 \Phi\,\, \del_\a{}^{\dot\a}\del_\beta{}^{\dot\beta}
\Lambda_{\dot\a\dot\beta}\,\, \del^{\a\dot\gamma}\del^{\beta\dot\delta}
\Lambda_{\dot\gamma\dot\delta}  +
\del_\a{}^{\dot\a}\del_\beta{}^{\dot\beta}
\Lambda_{\dot\a\dot\beta}\,\, \del^{\a\dot\gamma} \lambda_{\dot\gamma}\,\,
\del^{\beta\dot\delta}\zeta_{\dot\delta}\ ,\nonumber
\end{eqnarray}
where we define $\del_\a \equiv \del_\a{}^{\dot1}$ and $\del_{\tilde\a}
\equiv\del_\a{}^{\dot2}$.    The terms in this Lagrangian that involve only
the $\Phi$ field were first derived by Ooguri and Vafa \cite{OV} for
the closed $N=2$ string with the complex structure on the $(2,2)$ spacetime.
This part of the Lagrangian describes self-dual gravity in $2+2$ dimensions.
The metric is described by $g_{\a\tilde\beta}=g_{\tilde\beta\a} = \del_\a\,
\del_\tb\, K$, where $K$ is the K\"ahler potential $K= X_\ada\, X^\ada +
\phi$.  The line element is given by $ds^2=g_{\a \tilde\beta}\,
dX^{\a\dot1}\,dX^{\tilde\beta\dot2}$.   The reason for introducing the
tilded indices is that we are describing the $(2,2)$ spacetime in terms of a
real structure rather than the complex structure. Here, one uses double
numbers of the form $x+ e y$, where $e^2=1$, rather than complex numbers
\cite{bgppr}.  The tilde denotes the conjugation operation under which
$\tilde e = - e$.   Variations of the Lagrangian with respect to $\chi_\a$,
$\psi_\a$ and $F_{\a\beta}$ give the expressions of these fields in terms of
potentials $\lambda_{\dot\a}$, $\zeta_{\dot\a}$ and
$\Lambda_{\dot\a\dot\beta}$ respectively.

\section{Conclusions}

     In this paper, we have investigated the physical spectrum and
interactions of the $N=2$ string with a real structure in the $(2,2)$
spacetime by using BRST methods.  Several new features have come to light.
In the BSRT formalism, the ghosts for the fermionic currents have to be
bosonised, and since this is a non-local field redefinition, it has the
consequence that the physical spectrum can be altered by performing a
rotation of the two fermionic currents of the $N=2$ superconformal algebra.
 This rotation can map a physical state with finitely many terms in one
basis into a state with infinitely many terms in another basis.  The
allowable class of physical states and their conjugates has to be carefully
defined, and can lead to different cohomologies in different bases, as
discussed in section 4.  This can be illustrated by studying the structure
of the picture-changing operators.   In a generic basis for the fermionic
currents, the two picture-changing operators are invertible, and hence the
physical operators in different pictures can be identified. However, in
degenerate cases, for example the diagonal basis where the two fermionic
currents are $\pm$ eigenstates under the spin-1 current, the
picture-changing operators do not have inverses, and consequently the
physical spectrum is enlarged.

      We studied the physical spectrum in the off-diagonal basis.  There is
one scalar NS operator and one bosonic spinorial R operator.  In this
off-diagonal basis, there is no local operator of the form $\exp(\lambda
\int^z J^{\rm tot})$ that relates the physical operators in the NS and R
sectors.  Thus in the off-diagonal basis, the NS and R operators cannot be
identified.  This idea was reinforced by studying the interaction of the NS
and R operators.  There are two different types of three-point amplitudes,
namely $\langle {\rm NS, NS, NS}\rangle$ and $\langle {\rm R, R,
NS}\rangle$.  It is easy to show that at certain momentum values, the former
vanishes while the latter does not.

      In the diagonal basis, the descriptions are quite different, yet the
final conclusion is rather similar.  In this case, the NS and R operators can
be related by a local operator $\exp(\lambda \int^z J^{\rm tot})$ which
describes the moduli of the world-sheet $U(1)$ gauge field.  However, in the
diagonal basis, the picture-changing operators are not invertible, and the
physical operators at different pictures cannot be identified for all
on-shell momenta.  The consequence of this is that the three-point
interactions of all the physical operators involve three different types of
amplitudes.  These interactions, however, can be effectively described by
the interactions of a scalar NS operator and a spinorial R operator, which
is precisely the case for the off-diagonal basis.

     The differences between the NS and R operators are highlighted in the
real structure for spacetime that we consider in this paper.   The
computations for both the diagonal basis and the off-diagonal basis are
equally valid if one chooses the complex structure for spacetime.   In the
complex structure, there is a further constraint that the momentum and its
complex conjugate are not independent variables.  In particular, the
momentum and its complex conjugate vanish simultaneously.  In this case, the
three-point amplitudes $\langle {\rm NS, NS, NS}\rangle$ and $\langle {\rm
R, R, NS}\rangle$ can be viewed as equivalent since they are proportional to
each other with the proportionality never singular.   Thus in view of the
result we obtained in the real structure for spacetime, the identification
of the NS and R operators that was discussed for the complex structure is a
consistent choice, rather than a necessity imposed by the symmetry of the
theory.

     Finally, we comment on the possibility of spacetime supersymmetry.  All
the physical operators we have discussed in this paper are bosonic, and
reside in the (NS-NS) and (R-R) sectors with respect to the two fermionic
currents.   There are also (NS-R) and (R-NS) sectors \cite{gomis}, in which
physical operators with fermionic statistics would reside.   Certain
physical states in these two sectors were constructed in Ref.~\cite{BKL};
however, they involve twist operators in spacetime, which break the
spacetime structure.  This means that these states do not exist for general
on-shell momenta, but only in special cases where certain components vanish.
   It may be of interest to investigate the resulting two-dimensional theory
in more detail.

\section*{\bf Acknowledgements}

    We are grateful to SISSA, Trieste for hospitality during the course of
this work.  We thank the referee, N. Berkovits, E. Gava, R. Iengo, S.V.
Ketov, O. Lechtenfeld, K. Narain and A. Schwimmer for useful discussions.

\section*{\bf Appendix}

     In this appendix, we shall calculate the inverses $(\widetilde
Z^1)^{-1}$ and $(\widetilde Z^2)^{-1}$ of the picture-changing operators for
the $N=2$ string in a generic off-diagonal basis.   We shall first derive an
expression for the operator $(\widetilde Z^1)^{-1}$.  In section 3, we
showed that the leading term of $(\widetilde Z^1)^{-1}$ is given by
({\ref{picleading}).  By itself, this is not annihilated by the BRST
operator (\ref{offdiagbrst}).  As in Ref.~\cite{BKL}, we can remedy this by
adding an infinite sequence of terms, of the form
\begin{equation}
B_n = c\, \del^n\xi^1\cdots\del\xi^1\, \del^{n-2}\eta_2\cdots\eta_2\,
e^{-(n+1)\phi_1 + (n-1)\phi_2}\ .\label{bnseq}
\end{equation}
Most terms in the BRST operator (\ref{offdiagbrst}) annihilate the $B_n$
operators.   The only terms that do not are the following
\begin{eqnarray}
Q_1&\equiv&  x_2\, b\, \del\eta_1\, \eta_1\, e^{2\phi_1} -
{2x_2 \over 1-x_1\, x_2}\, \gamma\, \eta_1\, \del\xi^2\,
e^{\phi_1 -\phi_2}\ ,\nonumber\\
Q_2&\equiv& (1+x_1\, x_2) \Big( b\, \eta_1\, \eta_2 e^{\phi_1 + \phi_2}
+ {1\over 1-x_1\, x_2}\, \gamma (\del\phi_1 -\del\phi_2)\Big)\ ,
\label{q123}\\
Q_3&\equiv& x_1\, b\,\del\eta_2\, \eta_2\, e^{2\phi_2} +
  {2x_1 \over 1- x_1\, x_2}\, \gamma\, \del\xi^1\, \eta_2\,
e^{-\phi_1 + \phi_2}\ .\nonumber
\end{eqnarray}
Acting with these on $B_n$, we find
\begin{equation}
Q_1\, B_n = n!(n-1)!\, x_2, \widetilde B_{n-2}\ , \quad
Q_2\, B_n = -n (1+x_1\, x_2) \widetilde B_{n-1}\ ,\quad
Q_3\, B_n = {x_1 \over n!(n-1)!}\, \widetilde B_n\ ,
\label{q123bn}
\end{equation}
where $\widetilde B_n \equiv \del^n\xi^1\cdots\del\xi^1\, \del^n\eta_2\cdots
\eta_2\, e^{-(n+1)\phi_1 + (n+1)\phi_2} - 2\gamma\, B_{n+1}/(1-x_1\, x_2)$.
Here $B_n$ and $\widetilde B_n$ are defined to be zero when $n$ is negative.
{}From (\ref{q123bn}), it follows that
\begin{equation}
b_n\, Q_3\, B_n +b_{n+1}\, Q_2\, B_{n+1} + b_{n+2}\, Q_1\, B_{n+2} = 0\ ,
\end{equation}
where the coefficients $b_n$ satisfy the recursion relation
\begin{equation}
(n+1)!(n+2)!\, x_2\, b_{n+2} - (n+1) (1+ x_1\, x_2)\, b_{n+1} +
{x_1 \over n!(n-1)!}\, b_n = 0\ .
\end{equation}
This recursion relation can be easily solved to give
\begin{equation}
b_n = {1\over n} \prod^{n-1}_{j=1} (j!)^{-2}\, { 1- (x_1 x_2)^n \over
1 - x_1\, x_2}\, b_1 \ .
\end{equation}
Thus the inverse operator $(\widetilde Z^1)^{-1}$, which takes the form
\begin{equation}
(\widetilde Z^1)^{-1} = \sum_{n\ge 1} b_n\, B_n\label{z-1}
\end{equation}
with $b_1=1/x_2$, is annihilated by the BRST operator.   Similarly, the
inverse operator $(\widetilde Z^2)^{-1}$ is given by (\ref{z-1}),  with
indices 1 and 2 interchanged, and $b_1 = 1/x_1$.

\end{document}